\documentclass[utf8]{frontiersSCNS} 

\usepackage{url,hyperref,lineno,microtype,subcaption}
\usepackage[onehalfspacing]{setspace}
\usepackage{amsmath}
\usepackage{color} 

\def\keyFont{\fontsize{8}{11}\helveticabold }
\def\firstAuthorLast{Sawicki {et~al.}} 
\def\Authors{Jakub Sawicki\,$^{1,2,3,4,*}$, Lenz Hartmann$^{5}$, Rolf Bader$^{5}$ and Eckehard Schöll\,$^{1,4,6}$}
\def\Address{$^{1}$Potsdam Institute for Climate Impact Research, Telegrafenberg A 31, 14473 Potsdam, Germany \\
$^{2}$Institut f{\"u}r Musikp{\"a}dagogik, Universität der K{\"u}nste Berlin, Hardenbergstra\ss{}e 33, 10623 Berlin, Germany\\
$^{3}$Fachhochschule Nordwestschweiz FHNW, Leonhardsstrasse 6, 4009 Basel, Switzerland\\
$^{4}$Institut f{\"u}r Theoretische Physik, Technische Universit{\"a}t Berlin, Hardenbergstra\ss{}e 36, 10623 Berlin, Germany\\
$^{5}$Institute of Systematic Musicology, University of Hamburg, Neue Rabenstr. 13, 20354 Hamburg, Germany\\
$^{6}$Bernstein Center for Computational Neuroscience Berlin, Humboldt-Universit{\"a}t, Philippstra{\ss}e 13, 10115 Berlin, Germany
}
\def\corrAuthor{Jakub Sawicki}
\def\corrEmail{zergon@gmx.net}

\begin{document}
\onecolumn
\firstpage{1}

\title[Modelling the perception of music in brain network dynamics]{Modelling the perception of music in brain network dynamics} 

\author[\firstAuthorLast ]{\Authors} 
\address{} 
\correspondance{} 

\extraAuth{}

\maketitle

\begin{abstract}


We analyze the influence of music in a network of FitzHugh-Nagumo oscillators with empirical structural connectivity measured in healthy human subjects. We report an increase of coherence between the global dynamics in our network and the input signal induced by a specific music song. We show that the level of coherence depends crucially on the frequency band. We compare our results with experimental data, which also describe global neural synchronization between different brain regions in the gamma-band range and its increase just before transitions between different parts of the musical form (musical high-level events). The results also suggest a separation in musical form-related brain synchronization between high brain frequencies, associated with neocortical activity, and low frequencies in the range of dance movements, associated with interactivity between cortical and subcortical regions.
\tiny
 \keyFont{ \section{Keywords:} synchronization, coupled oscillators, neural network dynamics, pattern formation: activity and anatomic, external driven} 
\end{abstract}

\section{Introduction}

Dealing with the dynamics of neural networks, one repeatedly encounters the phenomenon of synchronization. In the brain, a high degree of synchronization is related to (slow-wave) sleep \citep{STE93b,RAT00} or transitions from wakefulness to sleep \citep{SCH08o,MOR12}. Recently, partial synchronization \cite{SCH21} has become a reference point for the explanation of the first-night effect \citep{TAM16} and unihemispheric sleep \citep{RAT00,RAT16,MAS16,RAM19}. Furthermore, synchronized dynamics plays an integral role in the dynamics of epileptic seizures \citep{GER20}, where the synchronization of a part of the brain causes dangerous consequences for the persons concerned. By contrast, synchronization is also used to explain brain processes serving the development of syntax and its perception \citep{KOE13,LAR15a,BAD20}. Generally, synchronization theory is of great importance for the analysis and understanding of musical acoustics and music psychology \citep{JOR94,BAD13,SAW18a,HOU20,SHA20}. Although the neurophysiological processes involved in listening to music are still being researched, it is believed that some degree of synchrony can be observed in listening to music and building expectations. Event-related potentials, measured by electroencephalography (EEG) of participants while listening to music, show synchronized dynamics between different brain regions \citep{HAR14, HAR20a}. These studies indicate that the increase of synchronization represents musical large-scale form perception. In more detail, the authors observed the strongest brain activity at frequencies in the beta-band (about $13-30$ Hz) and in the gamma-band (about $30-120$ Hz): In these bands, the increase and decrease of synchronization are following the large-scale form of the listened music in a coherent way. Moreover, it has been observed that areas of the whole brain are involved in neural dynamics during perception \citep{BAD20}. On the other hand, the general influence of sound on empirical brain networks has been investigated recently \citep{SAW21a}. It has been shown that an external sound source, which is connected to the auditory cortex of the human brain, induces partial synchronization patterns. Nevertheless, this study has neglected the complexity of music and its distinct effects in different frequency bands within the brain oscillations. There are a variety of recognized modeling approaches with respect to neural systems in general \citep{KAC15, BAS17, BAS18, PET18a, PET19} and related to music in particular \citep{FRI13}. In this paper, we model the spiking dynamics of the neurons by the paradigmatic FitzHugh-Nagumo model, and investigate possible coherence between the dynamics of the brain network and an external music source, which is connected to the auditory cortex of the human brain. Moreover, we present experimental data which we successfully reproduce numerically with the help of our network model, which combines simple node dynamics with complex network connectivities derived from empirical measurements.

An intriguing synchronization phenomenon in networks is relay (or remote) synchronization between layers which are not directly connected, and interact via an intermediate (relay) layer \citep{LEY18}. The simplest realization of such a system is a triplex network where a relay layer in the middle acts as a transmitter between the two outer layers. Remote synchronization, a regime where pairs of nodes synchronize despite their large distances on the network graph, has been shown to depend on the network symmetries \citep{BER12,NIC13,GAM13,ZHA17,ZHA17a}. Recently the notion of relay synchronization has been extended from completely synchronized states to partial synchronization patterns in the individual layers of a three-layer multiplex network. It has been shown that the three-layer structure of the network allows for (partial) synchronization of chimera states in the outer layers via the relay layer \citep{SAW18c,SAW18,SAW20,WIN19,DRA20,SAW21}. Going towards more realistic models, time-delay plays an important role in the modeling of the dynamics of complex networks. In brain networks, the communication speed is affected by the distance between regions and therefore a stimulus applied to one region needs time to reach a different region. In such delayed system, it is possible to predict if the effects of stimulation remain focal or spread globally \citep{MUL16}. More generally, time delays due to propagation over the white-matter tracts have been shown to organize the brain network synchronization dynamics for different types of oscillatory nodes \citep{PET19}. Within the scope of this paper, we focus on the requirements for a simple model to exhibit partial synchronization patterns, which have been experimentally observed \citep{HAR14, HAR20a}. Therefore, we defer the consideration of time delays for now.

This article is organized as follows. In Sect.\,2 we discuss the transformation of music to a neural input signal using a detailed cochlea model. In Sect.\,3 we introduce the neural network model based upon empirical connectivities with neural input to the auditory cortex generated by music. In Sect.\,4 we introduce some methods to characterize the neural output. Sect.\,5 presents the results of the computer simulations and discusses the dynamical scenarios. Sect.\,6 presents a comparison with experiments on human subjects, and Sect.\,7 finally concludes.

\section{From sound to neural spikes}

The transformation of sound into neural spikes is the subject of much current research \citep{TRI10, MIZ14, BAD15, BAD17, BAD18, GUO21}. Music, speech, or any sound enters through the outer and middle ear as sound pressure, then acting on the oval window of the cochlea. The movement of the oval window is then transferred to a pressure in the lymph liquid of the cochlea surrounding the basilar membrane, which again acts on the basilar membrane, causing traveling waves there. Due to spatial differences in stiffness and damping on the membrane, sinusoidal waves with a single frequency show an increase in amplitude up to a point with maximum amplitude, the position of the so-called best-frequency, with a fast decay afterwards. Therefore, different positions on the basilar membrane represent different frequencies, making the cochlea a Fourier analyzer. The stereocilia on the basilar membrane at the position of respective best-frequency are then transferring the mechanical energy into neural spikes. The frequency distribution on the basilar membrane is logarithmic. Movements of neighboring frequencies lead to interactions, causing roughness perception up to a frequency band of a musical major third. These bands are called critical bands, and the basilar membrane consists of 24 such bands. The spikes leaving the respective bands are fed into the auditory pathway, consisting of several neural nuclei, where the nucleus cochlearis or the trapezoid body are the first two. The interaction between these neural nuclei is manifold with several feedback loops and binaural connections \citep{SCH11k} ending at the auditory cortex on both hemispheres. Still up to the A1 region of the auditory cortex, the critical bands are maintained, where neural connections of higher nuclei are connected to bands on the basilar membrane, which is called tonotopy. 

Many auditory features are present, extracted, or perceived already in this pathway, like sound localization, pitch, or timbre \citep{LYO96}, although research has not concluded on further processing in the cortex \citep{BAD21}. Music perception of larger temporal content, like song or sonata form, are not part of processing in the auditory pathway up to the cortex, as far as we know. Still the feedback loops within the pathway are both directions, up and down, afferent and efferent, so e.g. there is one connection down from the cortex to the cochlea with only one nucleus in between, tuning the basilar membrane tension through efferent nerves, according to cortex activity \citep{SCH11k}.

Up to now, no model of the whole auditory pathway exists on a detailed neural level. The model used in this paper therefore concentrates on main findings, i.e., the transition from sound to neural spikes, the tonotopy of neural connections up to the cortex, as well as partial synchronization of phases in the cochlea, which are also present as coincidence detection in the auditory pathway. A Finite-Difference Time Domain (FDTD) physical model of the cochlea is used \citep{BAD15}. The basilar membrane is about 3.5 cm long and only between 0.1 - 0.12 cm wide, so it is more a rod than a membrane. Therefore, the present model assumes a differential equation of a membrane like

\begin{equation}
\frac{K(x)}{\mu(x)}\frac{\partial^2 u}{\partial x^2} - d(x) \frac{\partial u}{\partial t} = \frac{\partial^2 u}{\partial t^2} + f(t)\ ,
\label{eq:BMD}
\end{equation}
with basilar membrane displacement $u$ along a one-dimensional axis x, basilar membrane stiffness $K(x) = 2 \times 10^9 e^{-3.4 x}\, {\text{dyn}} / {\text{cm}}^3$ changing along $x$, and linear mass density $\mu(x) = m / A(x)$ with mass $m$ over cross section $A$ again changing along the basilar membrane and $A(x) = 0.1\, {\text{cm}} \times (0.1\, {\text{cm}} + 0.02\, {\text{cm}} \times x / l)$ with basilar membrane length $l = 3.5$ cm taking into account the slight widening of the basilar membrane over its length. The boundary conditions of the basilar membrane are homogeneous Dirichlet boundary conditions which do not allow for displacements on the boundaries, but any derivative is allowed in accordance with the physiological conditions. Comparison between a membrane and a rod model shows no considerable differences, therefore a rod model is used. Here $d$ is damping, and $f(t)$ is the driving force of the lymph fluid which drives the basilar membrane.

To calculate the spikes omitted by the cochlea, the recording of the musical piece used is fed into the cochlea model. Here the amplitudes of the digital musical sound file are taken as sound pressures acting on the oval window of the cochlea and therefore immediately on the peri- and endolymph around the basilar membrane. As the speed of sound in the lymph ($\sim$~1500 m/s) is much larger than the speed of waves on the basilar membrane which is between $\sim$~100 m/s at the oval window and down to $\sim$~10 m/s at the helicotrema, an instantaneous action of the pressure at the oval window on the basilar membrane is reasonable and known as long-wave approximation \citep{DEB91}. This holds for frequencies up to $\sim$~4 kHz, where pitch perception stops and humans only hear a very high sound. This approximation is used in the model. It leads to the force $f(t)$ in Eq.\,\eqref{eq:BMD} which represents the amplitudes of the digital musical sound file acting instantaneously on all points of the basilar membrane at each time point respectively. It is interesting to see that the traveling wave on the basilar membrane is therefore not caused by an external input slowly traveling through the cochlea but by the intrinsic solution of the inhomogeneous differential equation of the basilar membrane driven by a periodic force over its whole length instantaneously.

Depending on the brain region, neurological measurements reveal different time scales \citep{SPI20}. In our work we choose 50 ms as a time integration step as this is consistent with a characteristic time scale in music as well as in visual perception. In music 50 ms correspond to the second integration time, below which two events cannot be distinguished one from another. This leads to a threshold of 20 Hz, above which musical pitches are perceived and below which adjacent events are heard as rhythms. In vision, 18-24 frames per second lead to a continuous visual perception, again corresponding to about 50 ms time intervals. Therefore, in terms of hearing and seeing, the brain seems to update perceptional input on a time-scale of 50 ms \citep{BAD13}. 

The transition between mechanical displacement and electrical spike is performed using two conditions according to literature \citep{HUB96}. A neural spike at one point X on the basilar membrane at time $\tau$ is excited if two conditions hold 
\begin{subequations}
\begin{align}
	&u(X,\tau) > u(X-1,\tau), u(X+1,\tau)\\
	&u(X,\tau) > u(X,\tau-1), u(X,\tau+1).
\end{align}
\label{eq:cond}
\end{subequations}
Condition (\ref{eq:cond}a) means a maximum shearing of two nervous fibers as a necessary condition to an opening of the ion channels at the fibers. This only happens with a positive slope, as only then the stereocilia are driven away from each other. With a negative slope the cilia are getting closer and therefore no stress appears at the tip links between them. This corresponds to the rectification process in gammatone filter banks. Condition (\ref{eq:cond}b) is a temporal maximum positive peak of the basilar membrane displacement. It is the temporal equivalent to the spatial condition of a maximum acceleration, where the tip link between the cell and its neighboring cells is most active.

To calculate the spikes omitted by the cochlea, the recording of the musical piece used is fed into the cochlea model. Therefore, the original piece, available as a digital recording of $44.1$ kHz sample rate (CD-Quality) is upsampled to $192$ kHz to meet Finite-Difference Time Domain (FDTD) stability criteria. The cochlea model is then run with a time-discretization step of $\Delta t$ = 1/192000 s. Each time when a neural spike appears, the time point, strength, and critical band of the spike is stored. Therefore, after processing, a time series $I(t)$ of all spikes leaving the cochlea is obtained.

\begin{figure}[tbp!]
\begin{center}
\includegraphics[width = .75\textwidth]{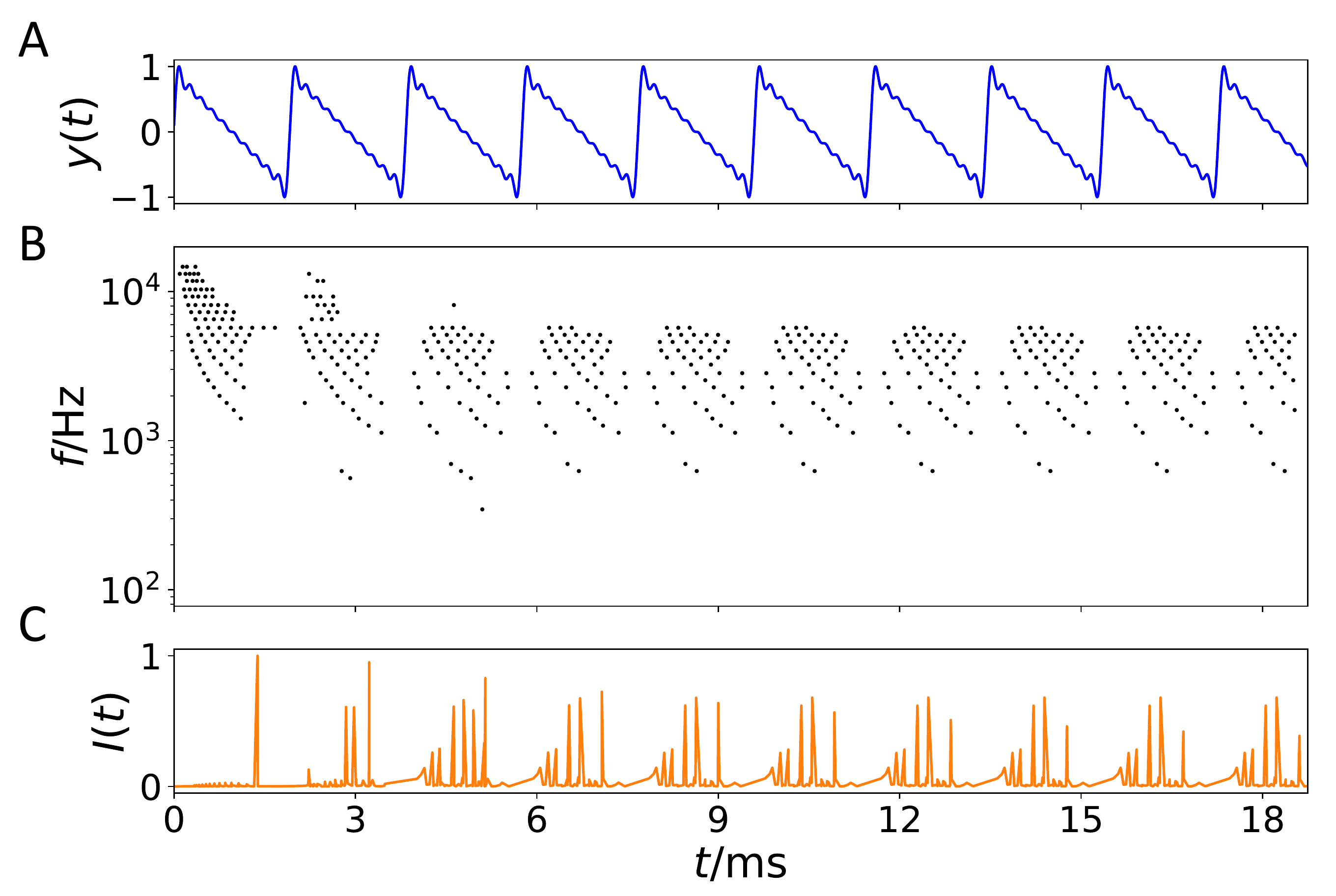}
\caption{Example of transformation of a sound wave into a spike pattern of the cochlea model. (A) Time series of an artificially generated tone complex $y(t)$ versus time $t$ in ms with $f_0 = 475$ Hz and ten partial tones (harmonics) with amplitudes $1/m$ where $m=1,2,3,\dots,10$. (B) Spikes (black dots) leaving the cochlea as calculated from the model \citep{BAD15}, where the vertical axis represents the cochlea position with best-frequency $f$ in Hz indicated, i.e., categorized into 24 so-called critical bands. (C) Time series $I(t)$ of the sum of all spike weights leaving the cochlea at a certain time $t$. Note that the first $5$ ms are transients.}\label{fig.1}
\end{center}
\end{figure}

Figure \ref{fig.1}A displays an example of an artificially generated so-called \textit{tone complex} with $f_0 = 475$ Hz and ten partial tones (harmonics) with amplitudes $1/m$ where $m=1,2,3,\dots,10$. The respective spike output of the basilar membrane model is shown in Fig.\,\ref{fig.1}B. Each time when the sound wave has a maximum amplitude, a pressure pulse is traveling over the basilar membrane, which emits electrical spikes at respective best-frequency positions on the membrane in accordance with the frequencies in the activating sound. As traveling waves on the membrane start at the basal end, next to the oval window, where high frequencies have their best-frequency location, and travel down the membrane towards the upper end, the helicotrema, where low frequencies are located, low frequencies show a time-delay with respect to higher frequencies. If the spikes of all critical bands are summed up for a certain point in time, a time series $I(t)$ of all neural spikes leaving the cochlea can be generated, as exemplarily shown in Fig.\,\ref{fig.1}C.

\section{Neural network model}

\begin{figure}[tbp!]
    \begin{center}
\includegraphics[width = .5\textwidth]{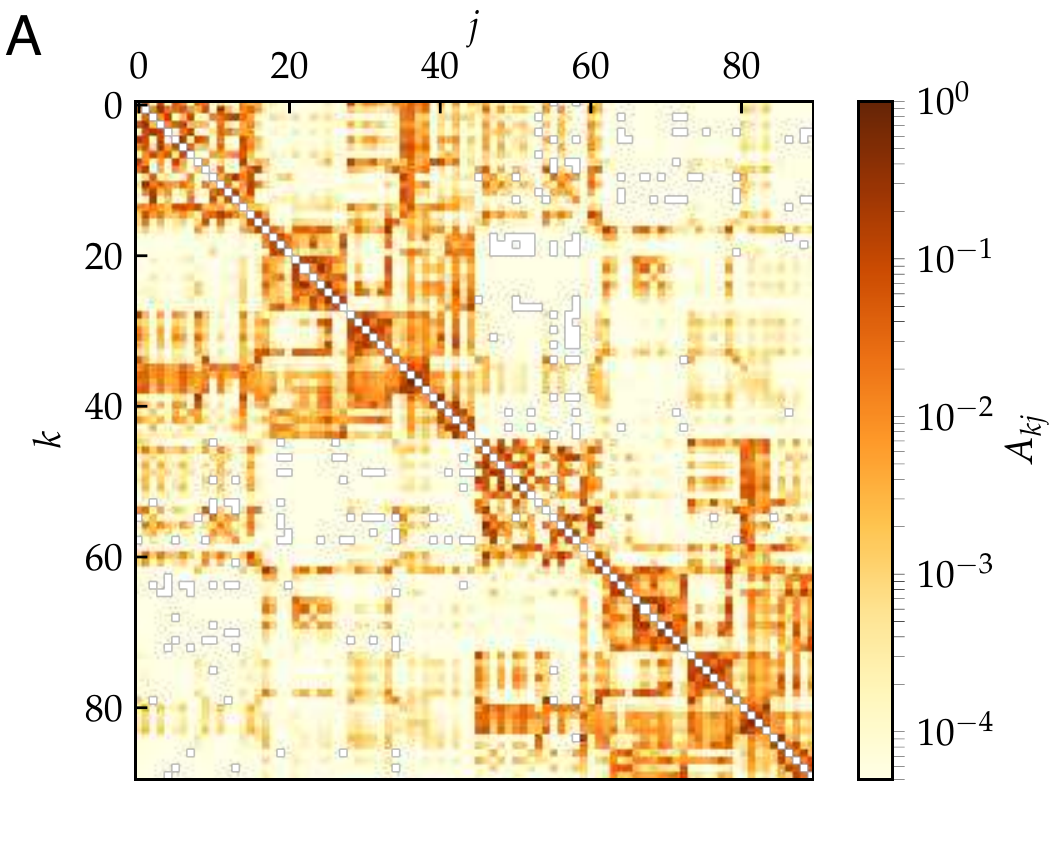}
\includegraphics[width = .8\textwidth]{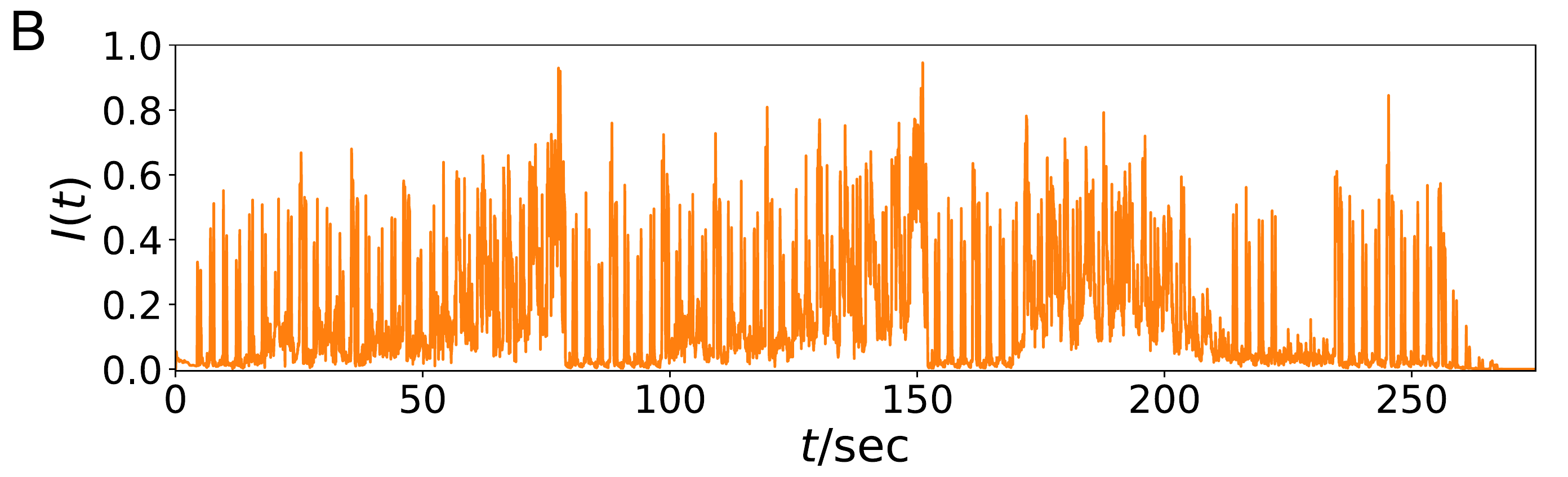}
    \caption{(A) Model for the hemispheric brain structure: Weighted adjacency matrix $A_{kj}$ of the averaged empirical structural brain network derived from twenty healthy human subjects by averaging over the coupling between two brain regions $k$ and $j$. The brain regions $k,j$ are taken from the Automated Anatomic Labeling atlas \citep{TZO02}, but re-labeled such that $k=1,...,45$ and $k=46,...,90$ correspond to the left and right hemisphere, respectively. After \citep{GER20}. (B) Time-series of the neural input signal $I(t)$ obtained from the music song \textit{One Mic} transformed by a method developed by Bader \citep{BAD20}. The song has a length of about 270 seconds and was released in 2002 by American rapper \textit{Nas}.}
    \label{fig.2}
    \end{center}
\end{figure} 

In this section, we introduce an empirical structural brain network as shown in Fig.\,\ref{fig.2}A where every region of interest is modeled by a single FitzHugh-Nagumo (FHN) oscillator. The weighted adjacency matrix $\mathbf{A} = \{A_{kj}\}$ of size $90 \times 90$, with node indices $k, j \in N = \{1,2,...,90\}$ was obtained from averaged diffusion-weighted magnetic resonance imaging data measured in 20 healthy human subjects. For details of the measurement procedure including acquisition parameters, see \citep{MEL15}, for previous utilization of the structural networks to analyze chimera states see \citep{CHO18,RAM19,GER20,SCH21}. The data were analyzed using probabilistic tractography as implemented in the FMRIB Software Library, where FMRIB stands for Functional Magnetic Resonance Imaging of the Brain (www.fmrib.ox.ac.uk/fsl/). The anatomic network of the cortex and subcortex is measured using Diffusion Tensor Imaging (DTI) and subsequently divided into 90 predefined regions according to the Automated Anatomical Labeling (AAL) atlas \citep{TZO02}, see Table~\ref{tab:AAL}. Each node of the network corresponds to a brain region. Note that in contrast to the original AAL indexing, where sequential indices correspond to homologous brain regions, the indices in Fig.\,\ref{fig.2}A are rearranged such that $k \in N_L = \{1, 2, ... ,45\}$ corresponds to left and $k \in N_R = \{46, ... ,90\}$ to the right hemisphere. Thereby the hemispheric structure of the brain, i.e., stronger intra-hemispheric coupling compared to inter-hemispheric coupling, is highlighted (Fig.\,\ref{fig.2}A). 

\begin{table}[tbp!]
\centering
\begin{tabular}{|c|c|c|} 
   \hline
   Label L/R & Region &Lobe \\
   \hline
   \hline
   1/46  & Precentral &  Central region\\
   2/47  & Frontal Sup &  Frontal lobe\\
   3/48  & Frontal Sup Orb & Frontal lobe\\
   4/49  & Frontal Mid &  Frontal lobe\\
   5/50  & Frontal Mid Orb & Frontal lobe\\
   6/51  & Frontal Inf Oper &Frontal lobe \\
   7/52  & Frontal Inf Tri  &  Frontal lobe\\
   8/53  & Frontal Inf Orb &  Frontal lobe\\
   9/54  & Rolandic Oper &  Central Region \\
   10/55 & Supp Motor Area &  Frontal lobe\\
   11/56 & Olfactory & Frontal lobe\\
   12/57 & Frontal Sup Medial &  Frontal lobe\\
   13/58 & Frontal Med Orb  &  Frontal lobe\\
   14/59 & Rectus & Frontal lobe\\
   15/60 & Insula  & Insula\\
   16/61 & Cingulum Ant  & Limbic lobe\\
   17/62 & Cingulum Mid & Limbic lobe\\
   18/63 & Cingulum Post & Limbic lobe\\
   19/64 & Hippocampus  & Limbic lobe \\
   20/65 & ParaHippocampal & Limbic lobe\\
   21/66 & Amygdala & Sub cort. gray nuc.\\
   22/67 & Calcarine  & Occipital lobe\\
   23/68 & Cuneus  & Occipital lobe\\
   24/69 & Lingual  & Occipital lobe\\
   25/70 & Occipital Sup & Occipital lobe \\
   26/71 & Occipital Mid  & Occipital lobe\\
   27/72 & Occipital Inf  & Occipital lobe \\
   28/73 & Fusiform   & Occipital lobe\\
   29/74 & Postcentral & Central region\\
   30/75 & Parietal Sup & Parietal lobe \\
   31/76 & Parietal Inf  & Parietal lobe\\
   32/77 & Supramarginal & Parietal lobe \\
   33/78 & Angular  & Parietal lobe\\
   34/79 & Precuneus & Parietal lobe\\
   35/80 & Paracentral Lobule & Frontal lobe\\
   36/81 & Caudate & Sub cort. gray nuc.\\
   37/82 & Putamen &Sub cort. gray nuc.\\
   38/83 & Pallidum &  Sub cort. gray nuc.\\
   39/84 & Thalamus & Sub cort. gray nuc.\\
   40/85 & Heschl &  Temporal lobe\\
   41/86 & Temporal Sup & Temporal lobe\\
   42/87 & Temporal Pole Sup & Limbic lobe\\
   43/88 & Temporal Mid & Temporal lobe\\
   44/89 & Temporal Pole Mid & Limbic lobe\\
   45/90 & Temporal Inf & Temporal lobe \\ \hline
\end{tabular}	
\caption{Cortical and subcortical regions, according to the Automated Anatomical Labeling atlas (AAL). Note that the numbering of the brain regions is different from the original numbering \citep{TZO02}.}
\label{tab:AAL}
\end{table}

The structural connectivity matrices serve as a realistic input for modeling, rather than as exact information concerning the existence and strength of each connection in the human brain. The pipeline for constructing such connectivity information using diffusion tractography is known to face a range of challenges~\citep{SCH19d}. While some estimates of the strength and direction of structural connections from measurements of brain activity can in principle be attempted, the relation of these can vary dramatically with (experimentally unknown) parameters of the local dynamics and coupling function~\citep{HLI12}.  

The auditory cortex is the part of the temporal lobe that processes auditory information in humans.  It is a part of the auditory system, performing basic and higher functions in hearing and is located bilaterally, roughly at the upper sides of the temporal lobes, i.e., corresponding to the AAL indexing $k = 41,86$ (temporal sup L/R). The auditory cortex takes part in the spectrotemporal analysis of the input passed on from the ear. Figure\ref{fig.2}B displays the time-series of impulses which are supplied to the brain by means of the auditory cortex. These neural impulses were obtained by the method of Bader described in Sect.\,2 \citep{BAD15,BAD17,BAD18}. Here, in contrast to Fig.\,\ref{fig.1}, a real piece of music was used, namely the hip hop music song \textit{One Mic}, composed by the American rapper \textit{Nas} and released in 2002. During the transition from acoustic mechanical to electrical excitation within the cochlea, synchronization appears to improve perception of pitch, speech, or localization. The sampling rate of these impulses obtained by Bader's method is $f_s = 192$ kHz. 
Each node corresponding to a brain region is modeled by the FitzHugh-Nagumo (FHN) model with external stimulus, a paradigmatic model for neural spiking \citep{FIT61,NAG62,BAS18}. Note that while the FitzHugh-Nagumo model is a simplified model of a single neuron, it is also often used as a generic model for excitable media on a coarse-grained level \citep{CHE05e,CHE07a}. Thus the dynamics of the network reads:
\begin{subequations}
\begin{align}
\epsilon \dot{u}_k = &u_k - \frac{u_k^3}{3} - v_k \nonumber \\
                    &+ \sigma \sum_{j \in N_\text{H}} A_{kj} \left[ B_{uu}(u_j - u_k) + B_{uv}(v_j - v_k) \right]  \\
                    &+ \varsigma \sum_{j \notin N_\text{H}} A_{kj} \left[ B_{uu}(u_j - u_k) + B_{uv}(v_j - v_k) \right], \nonumber \\
                    & +C_k I(t) \nonumber\\
\dot{v}_k = &u_k + a \nonumber \\
 & + \sigma \sum_{j \in N_\text{H}} A_{kj} \left[ B_{vu}(u_j - u_k) + B_{vv}(v_j - v_k) \right]  \\
 & + \varsigma \sum_{j \notin N_\text{H} } A_{kj} \left[ B_{vu}(u_j - u_k) + B_{vv}(v_j - v_k) \right], \nonumber
\end{align}
\label{eq.1}
\end{subequations}
with $k \in N_\text{H}$ where $N_\text{H}$ denotes either the set of nodes $k$ belonging to the left ($N_L$) or the right ($N_R$) hemisphere. Parameter $\epsilon = 0.05$ describes the timescale separation between the fast activator variable (neuron membrane potential) $u$ and the slow inhibitor (recovery variable) $v$ \citep{FIT61}. Depending on the threshold parameter $a$, the FHN model may exhibit excitable behavior ($\left| a \right| > 1$) or self-sustained oscillations ($\left| a \right| < 1$). We use the FHN model in the oscillatory regime and thus fix the threshold parameter at $a=0.5$ sufficiently far from the Hopf bifurcation point. 
The coupling within the hemispheres is given by the coupling strength $\sigma$ while the coupling between the hemispheres is given by the inter-hemispheric coupling strength $\varsigma$. As we are looking for partial synchronization patterns we fix $\sigma = 0.7$ and $\varsigma=0.15$ similar to numerical studies of synchronization phenomena during unihemispheric sleep \citep{RAM19} where partial synchronization patterns have been observed. The interaction scheme between nodes is characterized by a rotational coupling matrix:
\begin{equation}
\mathbf{B} = 
\begin{pmatrix}
B_{uu} & B_{uv} \\
B_{vu} & B_{vv}
\end{pmatrix}
=
\begin{pmatrix}
\text{cos}\phi & \text{sin}\phi \\
-\text{sin}\phi & \text{cos}\phi
\end{pmatrix},
\end{equation}
with coupling phase $\phi = \frac{\pi}{2} - 0.1$, causing primarily an activator-inhibitor cross-coupling. This particular scheme was shown to be crucial for the occurrence of partial synchronization patterns in ring topologies \citep{OME13} as it reduces the stability of the completely synchronized state. Also in the modeling of epileptic-seizure-related synchronization phenomena \citep{GER20}, where a part of the brain synchronizes, it turned out that such a cross-coupling is important. The subtle interplay of excitatory and inhibitory interaction is typical of the critical state at the edge of different dynamical regimes in which the brain operates \citep{MAS15a,SHI22}, and gives rise to partial synchronization patterns which are not found otherwise.

The external stimulus $I(t)$ describes the impulses evoked by the music piece \textit{One Mic} by \textit{Nas} and is applied to the brain areas $k=41,86$ associated with the auditory cortex, i.e., $C_k=1$ if $k=41$ or $86$ and zero otherwise. Since $I(t)$ is a time series which is calculated from a real piece of music, see Sect. 2, it has a physical dimension in seconds. On the other hand, the FitzHugh-Nagumo model uses dimensionless time. In order to compare our simulations with real data, we must transform the dimensionless time units of the FHN oscillator model to real time units by comparing the FHN oscillation period of a single FHN oscillator $T \approx 2.5$ (corresponding to dimensionless frequency $f_{\text{FHN}}\approx 0.4$) to the characteristic frequencies $n_b$ in Hz of an empirical time series. Depending upon the frequency band $n_b$ (in Hz) chosen, the simulation time is converted to real time by 1\,s $= 2.5 n_b$ simulation time units, or the simulated frequency (in Hz) is 
\begin{equation}
f_b=n_b/f_{\text{FHN}}.
\label{time_scaling}
\end{equation} 

\section{Methods}
We explore the dynamical behavior by calculating the mean phase velocity ${\omega_k = 2\pi M_k/\Delta T}$ for each node $k$, where $\Delta T$ denotes the time interval during which $M_k$ complete rotations are realized. Throughout the paper we denote the length of the input signal $I(t)$ as $\Delta T$. For all simulations we use initial conditions randomly distributed on the circle $u_k^2+v_k^2=4$ and a transient time of $t_{\text{trans}}=10\,000$ before the input signal $I(t)$ is supplied to the system. In case of an uncoupled system ($\sigma=0$), the mean phase velocity (or natural frequency) of each node is $\omega_k = \omega_{\text{FHN}} = 2\pi f_{\text{FHN}}\approx 2.51$. 

First, we introduce the spatially averaged mean phase velocity: 
\begin{equation}
{\overline\omega = \frac{1}{90}\sum_{k =1}^N \omega_k},    
\end{equation}
Thus $\overline\omega$ corresponds to the mean phase velocity averaged over the left and right hemisphere.
Second, the Kuramoto order parameter
\begin{equation}
{R(t) = \frac{1}{90} \left| \sum_{k =1}^N \text{exp}[i \theta_k(t)]\right|},
\end{equation}
is calculated by means of an abstract dynamical phase $\theta_k$ that can be obtained from the standard geometric phase ${\tilde{\phi}_k(t) = \text{arctan}(v_k/u_k)}$ by a transformation which yields constant phase velocity $\dot{\theta}_k$. For an uncoupled FHN oscillator the function $t(\tilde{\phi}_k)$ is calculated numerically, assigning a value of time $0<t(\tilde{\phi}_k)<T$ for every value of the geometric phase, where $T$ is the oscillation period. The dynamical phase is then defined as $\theta_k=2 \pi t(\tilde{\phi}_k)/T$, which yields $\dot{\theta}_k = \text{const}$. Thereby identical, uncoupled oscillators have a constant phase relation with respect to the dynamical phase. Fluctuations of the order parameter $R$ caused by the FHN model's slow-fast time scales are suppressed and a change in $R$ indeed reflects a change in the degree of synchronization. The Kuramoto order parameter may vary between 0 and 1, where $R=1$ corresponds to complete phase synchronization, and small values characterize spatially desynchronized states. 

Third, we introduce a new measure which specifies the coherence between the Kuramoto order parameter and the input signal by using the time average of the Kuramoto order parameter weighted with the input signal
\begin{equation}
\gamma = \frac{1}{\Delta T}\int_0^{\Delta T} R(t)I(t) \,\mathrm{d} t
\label{eq:gamma}
\end{equation}
to quantify the overlap of coherent episodes ($R$ large) with large input signals, averaged over time. The coherence $\gamma$ is maximum if the synchronization is large whenever the signal is large. It is small if the overall synchronization is low, or if the modulation of the synchronization in time is not in phase with the modulation of the input signal amplitude. For $\gamma=0$ the Kuramoto order parameter and the input signal do not overlap at any time point. An increased value of $\gamma \in [0,1]$ means increased overlap  between the Kuramoto order parameter and the input signal. The motivation for introducing the measure $\gamma$ lies in the fact that in the human brain the increase and decrease of synchronization follows the large-scale form of the listened music in a coherent way \citep{HAR14, HAR20a}.

The input signal $I(t)$ is obtained from the original music song \textit{One Mic} by the cochlea model described in Sect.\,2 (see Fig.\,\ref{fig.1}). The song has a length of about 4.5 minutes and the sampling rate of the obtained input signal is given by $f_s = 192$ kHz. Sampling is the reduction of a continuous-time signal to a discrete-time signal, e.g., the conversion of a sound wave (a continuous signal) to a sequence of samples (a discrete-time signal). The sampling rate $f_s$ is then the average number of samples obtained in one second. According to the Nyquist criterion, the frequency information of $I(t)$ is then band-limited to $f_b < \frac{1}{2}f_s$. 

The local dynamics of our model, the FitzHugh-Nagumo oscillator, has no explicit time scale. Its intrinsic angular frequency is dimensionless and given by $\omega_k = \omega_{\text{FHN}}= 2 \pi f_{\text{FHN}}\approx 2.51$. To include the time signal $I(t)$ correctly in our dimensionless model, we scale the dimensional neural input obtained in Sect.\,2 by the frequency $n_b$ (in Hz), or equivalently, convert the dimensionless output of our simulation into physical dimensions $f_b$ according to Eq.\,\eqref{time_scaling}. This parameter $n_b$ effectively removes the time scale from the input, but on the other hand, it can also be seen as creating a link between our dimensionless model and the input signal $I(t)$.


\section{Frequency bands and coherence}

\begin{figure}
\centering
\includegraphics[width = .8\textwidth]{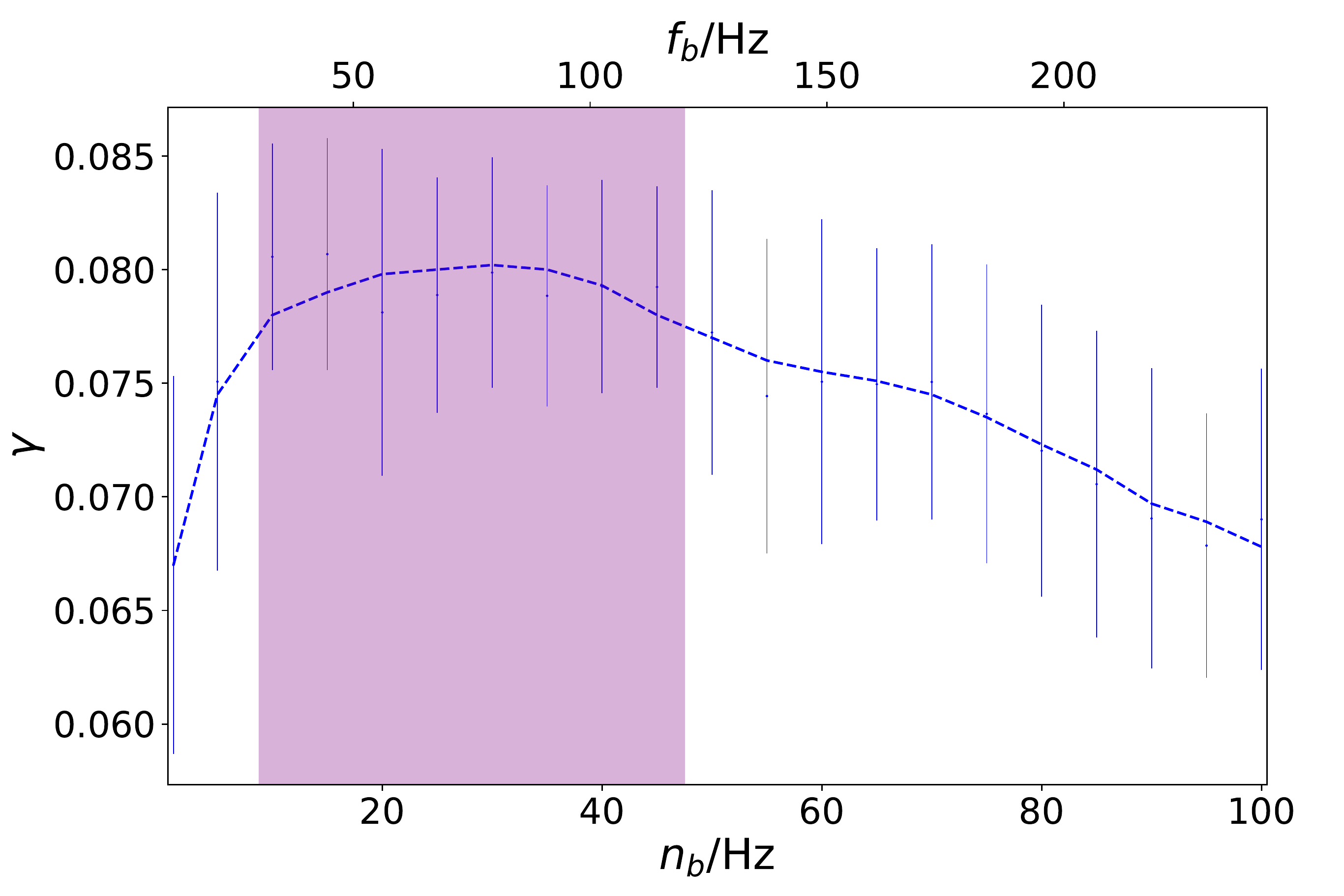}
    \caption{Coherence between network dynamics and external stimulus: coherence measure $\gamma$ in dependence on the characteristic music frequency $n_b$ (in Hz). The labeling on the upper x-axis denotes the corresponding frequency $f_b = n_b/f_{\text{FHN}}$ in the brain, where $f_{\text{FHN}}\approx 0.4$ is the dimensionless frequency of the FHN model, and the purple shaded region indicates the gamma-band ($f_b \approx 30-120$ Hz). The vertical bars indicate the standard deviation of the coherence measure for an ensemble of 200 simulations. The dashed line is obtained by a Savitzky–Golay filter. Other parameters are given by $\sigma=0.7$, $\varsigma=0.15$, $\epsilon = 0.05$, $a=0.5$, and $\phi = \frac{\pi}{2} - 0.1$.}
    \label{fig.3}
\end{figure}

Next, we investigate dynamical scenarios emerging from an external stimulus in the auditory cortices of both hemispheres ($k=41,86$). In order to compare our simulations with the empirical analysis of the influence of music upon the brain \citep{HAR14, HAR20a}, we may choose different frequency bands $n_b$, and hence a different scaling of the time in the external stimulus. This can be visualized by plotting the coherence measure $\gamma$ in dependence on the characteristic frequency $n_b$ (in Hz), see Fig.\,\ref{fig.3}. We find a strong non-monotonic behavior of $\gamma (n_b)$ and it turns out that by taking the frequency band $n_b$ of the external stimulus as a control parameter, one can change the level of coherence between the system dynamics and the external stimulus. Although the standard deviation of the coherence measure is relatively large for an ensemble size of 200 simulations (indicated by the vertical bars), we find a pronounced maximum of the coherence $\gamma$ for $n_b= 12-48$ Hz corresponding to the gamma-band of brain waves ($f_b \approx 30-120$ Hz) shown in Fig.\,\ref{fig.3} by purple shading. This means that for that frequency $n_b$ the level of synchronization follows the external signal most closely. It is in agreement with what has been observed in empirical brain analysis of the perception of music \citep{HAR14, HAR20a}. 

\begin{figure}
\centering
\includegraphics[width = .9\textwidth]{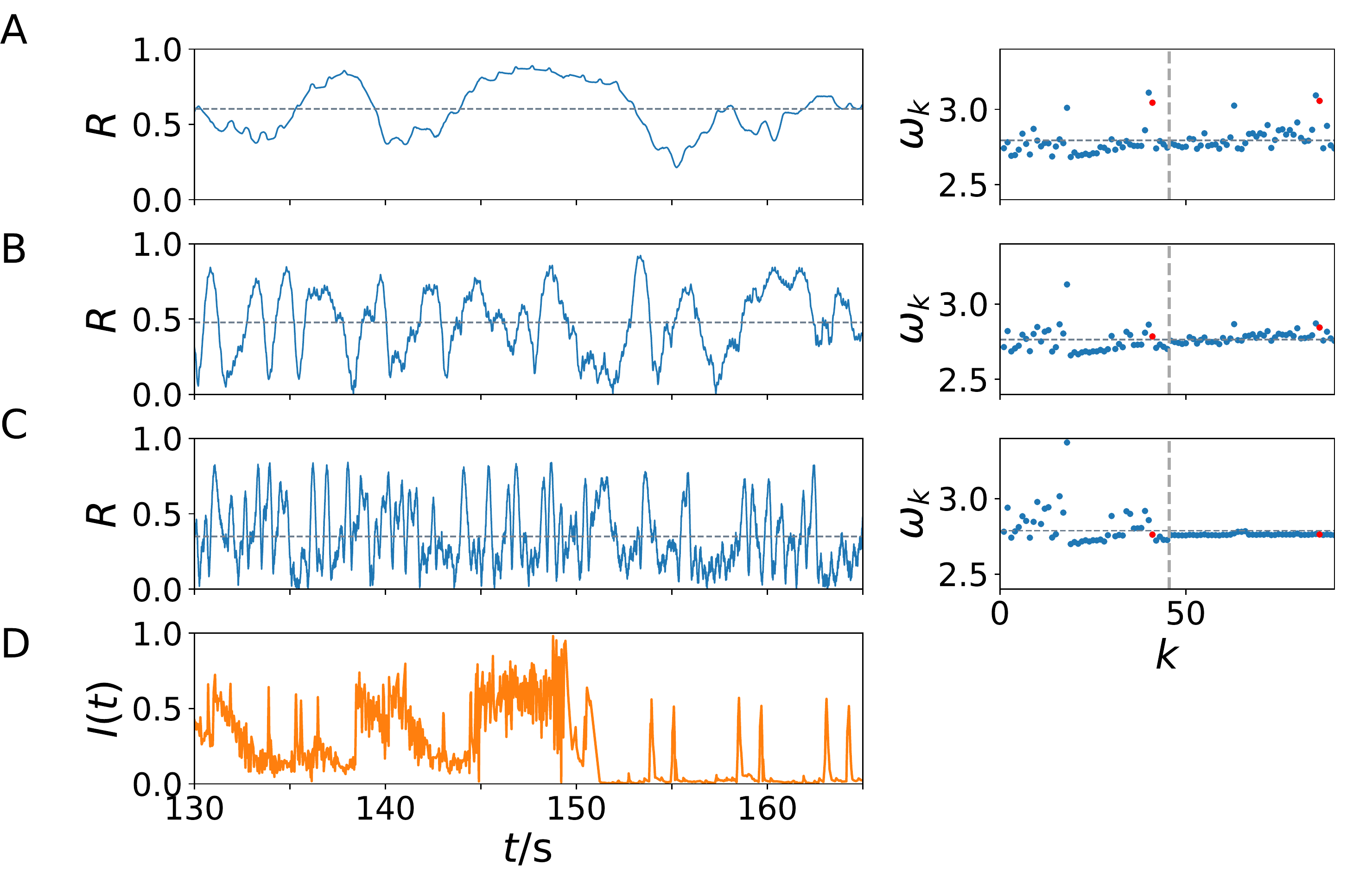}
    \caption{Dynamical scenarios: network dynamics for low and high values of coherence $\gamma$. Kuramoto order parameter $R$ versus time in s (left column) and dimensionless mean phase velocity profile $\omega_k= 2\pi f_k$ versus $k$ (right column) for increasing values of the frequency $n_b$ of the external stimulus $I(t)$ (A) $n_b=5$ Hz, (B) $n_b=30$ Hz and (C) $n_b=90$ Hz. In panel (D) the corresponding external stimulus $I(t)$ is plotted, which is a blowup of a part of Fig.\,\ref{fig.2}B. The vertical dashed line in the right column separates the left and right brain hemisphere; the red dots mark the nodes of the auditory regions ($k = 41, 86$). The horizontal grey dotted line indicates the temporal average of the Kuramoto order parameter $R$ in the left column, and the spatial average of the mean-field frequency $\overline\omega$ in the right column. Other parameters are as in Fig.\,\ref{fig.3}.}
    \label{fig.4}
\end{figure}

Figure \ref{fig.4}A-C depicts the details of the change of the time series of the Kuramoto order parameter $R(t)$ with increasing values of the frequency band $n_b$ of the external stimulus $I(t)$, which is shown in Fig.\,\ref{fig.4}D. It represents a part of the neural input signal $I(t)$ constructed from the music song \textit{One Mic} and shown in Fig.\,\ref{fig.2}B. We take a closer look at the temporal evolution of $R$ and the mean phase velocities $\omega_k$ in the system for different values of $n_b$ chosen from three different regimes in Fig.\,\ref{fig.3}: With increasing value of $n_b$ in panels (A)-(C), the time scale of the simulated neural output in Hz changes from lower to higher frequencies $f_b$ which is also seen in the temporal fluctuations of $R(t)$. Furthermore we observe on the one hand an increasing amplitude of the temporal fluctuations of $R$. On the other hand, the temporal average of the Kuramoto order parameter $R$ decreases with increasing $n_b$, marked by a horizontal grey dotted line in the left column: While for a small value of $n_b=5$ Hz in Fig.\,\ref{fig.4}A the Kuramoto order parameter $R$ assumes rather large values, and small values $R<0.2$ are not reached, for high values of $n_b=90$ Hz in Fig.\,\ref{fig.4}C rather small values of $R$ are measured. This trend can be seen by means of the temporal average of the Kuramoto order parameter $R$. For $n_b=30$ Hz in Fig.\,\ref{fig.4}B, the temporal average of $R$ takes a value $\approx 0.5$ and the time evolution shows regular oscillations between low ($R<0.2$) and high values ($R>0.8$). This aspect will be further discussed in the next section, since it can also be observed in experiments. 

As shown in Fig.\,\ref{fig.3}, in the case of $n_b=30$ Hz the coherence $\gamma$ is maximum.  Even though a higher value of the temporal average of $R(t)$, as observed in Fig.\,\ref{fig.4}A for $n_b=5$, might imply a higher value of $\gamma$ according to Eq.\,\eqref{eq:gamma}, Fig.\,\ref{fig.4}B shows that it is more important that $R(t)$ and $I(t)$ show a similar temporal modulation, as in Fig.\,\ref{fig.4}B for $n_b=30$. Despite the averaging over 250 simulations over the whole simulation time in Fig.\,\ref{fig.3}, the time segment in Fig.\,\ref{fig.4}B shows such a similarity in the modulation: We can see simultaneous drops of $R(t)<0.1$ and $I(t)<0.1$ for example at $t \approx 138,140,150$, whereas the values in between are higher, even if they fluctuate. 

In the right column of Fig.\,\ref{fig.4} the dimensionless mean phase velocities $\omega_k$ of all nodes are plotted, the horizontal grey dotted line indicates the spatial average, i.e., the collective mean-field frequency $\overline\omega$, which does not change for different $n_b$ since it is determined by the intrinsic collective dynamics. In contrast, the node dynamics of the auditory regions ($k = 41, 86$), indicated by red dots, depends on $n_b$ since it receives the external input signal which has a higher frequency in dimensionless units if the time is scaled in larger units $1/n_b$.  For $n_b = 5$ Hz in Fig.\,\ref{fig.4}A, the mean phase velocity of the auditory cortex is higher compared to the spatial average of the collective mean-field frequency $\overline\omega$. For $n_b = 30$ Hz in Fig.\,\ref{fig.4}B, the mean phase velocity of the auditory cortex approaches $\overline\omega$ having a bigger impact on the dynamics of the whole system than in Fig.\,\ref{fig.4}A for $n_b = 5$ Hz. Remarkable is the fact of a dynamical asymmetry shown by the mean phase velocities in Fig.\,\ref{fig.4}C: 

While the nodes of the right hemisphere exhibit equal mean phase velocity, i.e., they are frequency synchronized, the left hemisphere remains desynchronized and exhibits on average faster dynamics. This may indicate that regardless of the input $I(t)$ the system can exhibit partial synchronization. Such behavior is similar to the dynamics of unihemispheric sleep studied in \citep{RAM19}, where no external input has been applied to the dynamical system. In such states one hemisphere is synchronized, whereas the other hemisphere is partially desynchronized.

\section{Comparison with experiments}

The coupling of oscillatory neural signals within the usual frequency bands has been thought to be a mechanism that is related to a broad range of perceptual, sensorimotor, and cognitive processes, such as Gestalt perception and binding \citep{GRA89c,TAL95,KEI99,TAL99,ROD99a,ENG01b,ENG01c}, timing and expectation \citep{BUH05,BUH09}, attention \citep{WOM07a,FRI09c,NIK13a}, consciousness \citep{BAA06,DEH11b,ENG16,OWE19}, or motor functions \citep{THA15} as well as in music perception \citep{BHA01,ZAN05}. According to \citep{ENG16}, oscillatory brain activity is usually clustered into several frequency bands: delta ($0.5–3.5$ Hz), theta ($4-7$ Hz), alpha ($8-12$ Hz), beta ($13-30$ Hz) and gamma ($>30$ Hz). Since the gamma-band is the ‘youngest’ frequency band which has become of interest (from about the late 1990s), the ranges and definitions vary from source to source. Here, we refer to the classification of \citep{FRE13a}, who speak of  a low gamma range for frequencies above $30$ Hz up to $60$ Hz, and high gamma for frequencies above $60$ Hz up to about $120$ Hz. For everything above $120$ Hz, we use the term ‘fast oscillations’ as employed by \citet[p.114]{BUZ06}. The gamma-band frequency range is of particular interest in the context of large-scale synchronization since it is thought to be a mechanism that integrates information from different parts of the cortex.

\begin{figure}
\centering
\includegraphics[width = .9\textwidth]{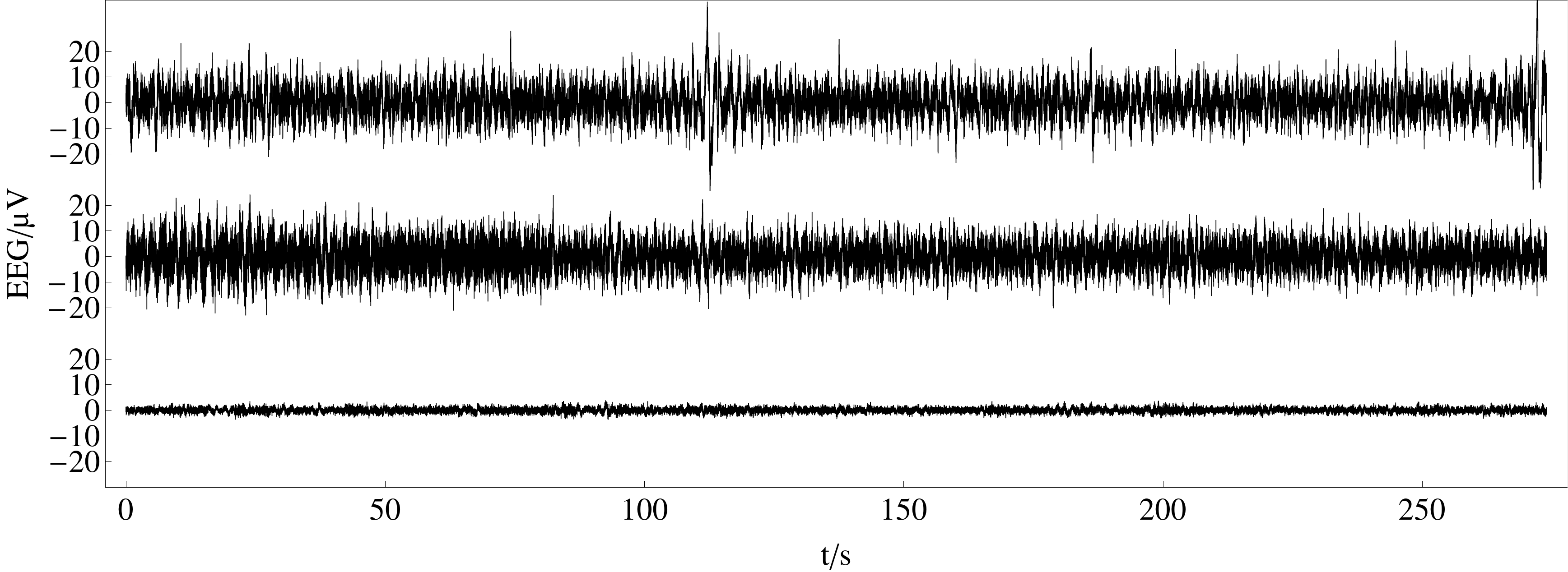}
    \caption{Recorded and averaged electroencephalogram (EEG) data: top and middle plot show recorded EEG time series after pre-processing for one electrode (Fp1) from two different participants. The bottom plot shows the time series of the same electrode averaged over 25 subjects and 3 trials.}
    \label{fig.5}
\end{figure}

The musical form as the hierarchically highest level of musical structure and its perception is related to some of the mentioned processes above \citep{LER90,HAR20a}. Perceptually, notes, bars, and phrases are grouped and integrated into a high-level part of the form by the Gestalt laws \citep{LEM97a,DEU13,NEU13,DEL14}. The contrast of the form’s parts, such as the concatenation of verse and chorus in a song, the sonata form of classical music, or the continuous night-long tension build-up and decay in Techno, House or Electronic Dance Music, characterize the musical form and the learned knowledge about the underlying structures leads to the build-up of expectation and their fulfillment as well as to modulated attention. On an emotional level, this can be expressed in the terms of tension and relaxation \citep{KOE14,LEH15c}. Also, the transition from “potential energy” (expectations) into “kinetic energy” (dancing) as proposed by \citep{KUR31} can be related to the processing of musical form in the sense of entrainment of neurons in the motor cortex by neurons from the auditory cortex \citep{THA15}.

The characteristic of contrasting parts can be revealed not only by music analysis using pen and paper but also by different computational methods by the music information retrieval discipline, like the amplitude of a piece of music that corresponds to the subjective perception of loudness. Also other properties of the stimulus, such as the spectral centroid that corresponds to the perceived brightness of a sound, or the fractal correlation dimension \citep{GRA83,GRA83a} corresponding to the perceived density and thereby representing the complexity of a piece of music, are drivers of the musical form \citep{BAD13,HAR20a,BAD21,LIN21b,BAD21a}.

Based on the aforementioned correlations between the processes associated with the perception of musical form and neural synchronization we expect the dynamics of neural synchronization to correspond to the dynamics of the amplitude of the stimulus. Again, the musical amplitude corresponds to perceived loudness, and is calculated as integration of energy over time intervals. Then synchronization between different brain regions is high when the amplitude of the musical piece is high, and synchronization is low when the amplitude of the piece is low. We expect such brain synchronization to be strong due to the prominence of the gamma-band in perception of musical parameters.

\begin{figure}
\centering
\includegraphics[width = .9\textwidth]{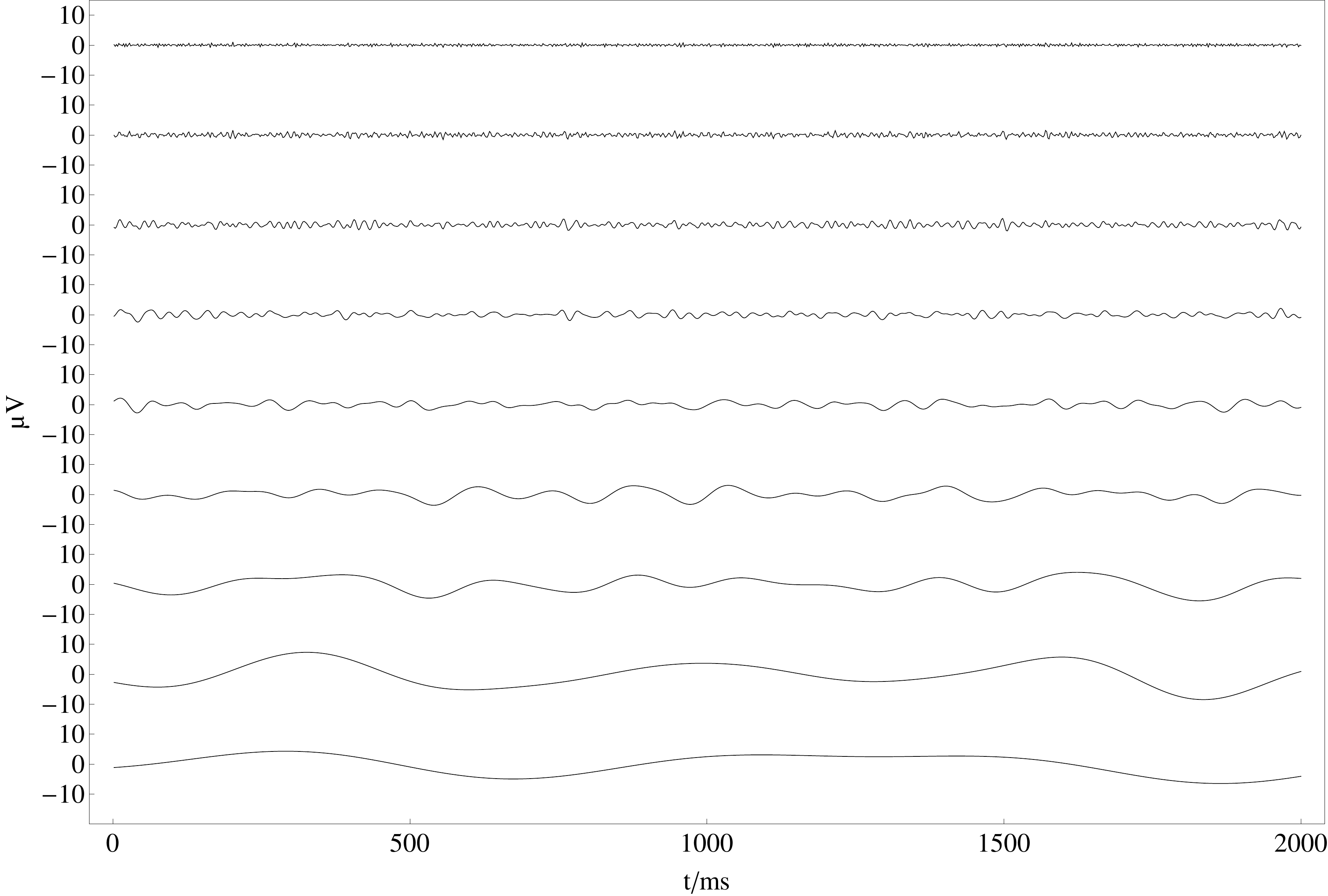}
    \caption{Nine frequency bands (FB) after wavelet transformation: Result of the continuous wavelet transform for the first two seconds of the averaged time series in Fig.\,\ref{fig.5}. From top to bottom frequency bands correspond to FB 1: $125 - 250$ Hz, FB 2: $62.5 - 125$ Hz, FB 3: $31.25 - 62.5$ Hz, FB 4: $15.63 - 31.25$ Hz, FB 5: $7.81 - 15.63$ Hz, FB 6: $3.91 - 7.81$ Hz, FB 7: $1.95 - 3.91$ Hz, FB 8: $0.98 - 1.95$ Hz, FB 9: $0.49 – 0.98$ Hz.}
    \label{fig.6}
\end{figure}

In an experiment, we have recorded the electroencephalogram (EEG) from human scalps to examine the perception of music large-scale form (Fig.~\ref{fig.5})\footnote{We have taken into account the usual guidelines regarding ethical procedure (informed consent). The subjects were mainly found and recruited through the Institute of Systematic Musicology Hamburg and had instrumental lessons on at least one instrument (mean duration $10.0$ years, standard deviation $4.6$ years) or corresponding experience as DJ. They participated in accordance with local ethics committee guidelines.}. 25 musically skilled subjects listened to the song \textit{One Mic} from the artist \textit{Nas} three times each. The song was released in 2001 on his Album \textit{Stillmatic} on Columbia Records. The electroencephalogram (EEG) signals were recorded with a sample rate of $500$ Hz from 32 electrodes, positioned following the 10-20 method of placement \citep{JAS58}. After artifact correction, recorded data for each channel has been averaged over subjects and trials to obtain a grand average of 75 trials for each channel to increase the signal to noise ratio and enhance event-related potentials. This type of averaging reveals \textit{evoked potentials} (in contrast to \textit{induced potentials}) and is related to the presented stimulus in a classical event-related potential manner \citep{TAL96, TAL99, ZAN05}. Since the positions of the electrodes do not differ over measurement time, the differences in correlation strength between different electrodes cannot be explained by spurious synchrony \citep{HOL77, KAY06, BHA18}. For a more detailed description of the experimental procedure, technical details and pre-processing, see \citep{HAR20a}.

In Fig.\,\ref{fig.6}, all channels have been decomposed into nine independent frequency bands that correspond approximately to the frequency bands mentioned above by using a continuous wavelet transformation with a Mexican Hat wavelet \citep{FRE13a}. In contrast to a bandpass filter with a subsequent Hilbert transform, using a Mexican hat wavelet for filtering is fast and efficient since one can decompose the recorded EEG data into the desired frequency-bands in one step by defining the number of octaves. The continuous wavelet transform of a uniformly sampled sequence $\{x_1,x_2,\dots\,x_n\} = \{x(t_0),x(t_0+\Delta t),\dots,x(t_0+(n-1) \Delta t)\}$ is given by
\begin{equation}
  w(u,s) = \frac{1}{\sqrt{s}}\sum\limits_{k=1}^n x_k \,\psi\left(\frac{(k-u)\Delta t}{s}\right),
\end{equation}
where $s \in \mathbb{R}$ corresponds to the frequency of the EEG band and $u=1,\dots, n$ labels the wavelet coefficients with the number $n$ of analyzed sample points defining the time window of observation. As wavelet function $\psi$ a Mexican Hat wavelet is used, given by
\begin{equation}
  \psi(x) = \frac{-2}{\sqrt[4]{\pi}\sqrt{3\sigma}}\left( \frac{x^2}{\sigma^2} -1\right) \exp\left(-\frac{x^2}{2\sigma^2} \right),
\end{equation}
where $\sigma$ is the width of the wavelet. The EEG bands used align very well with a musical scale, where each higher band doubles the frequency of its respective lower band, corresponding to a musical octave. Please note that this relation might only be at chance, still it may also relate to the fact that all human senses relate physics to perception in a logarithmic way \citep{SCH18l}. It is therefore convenient to scale $s$ in the wavelet transform in the same mathematical way as an equal-tempered musical scale like $s_{\text{oct}} = \alpha \,2^{\,\text{oct}-1}$, where $\text{oct} \in \{1,2,\dots,9\}$ is the octave number related to the nine frequency bands shown in Fig.\,\ref{fig.6} and $\alpha$ is the smallest wavelet scale.

\begin{figure}
\centering
\includegraphics[width = .9\textwidth]{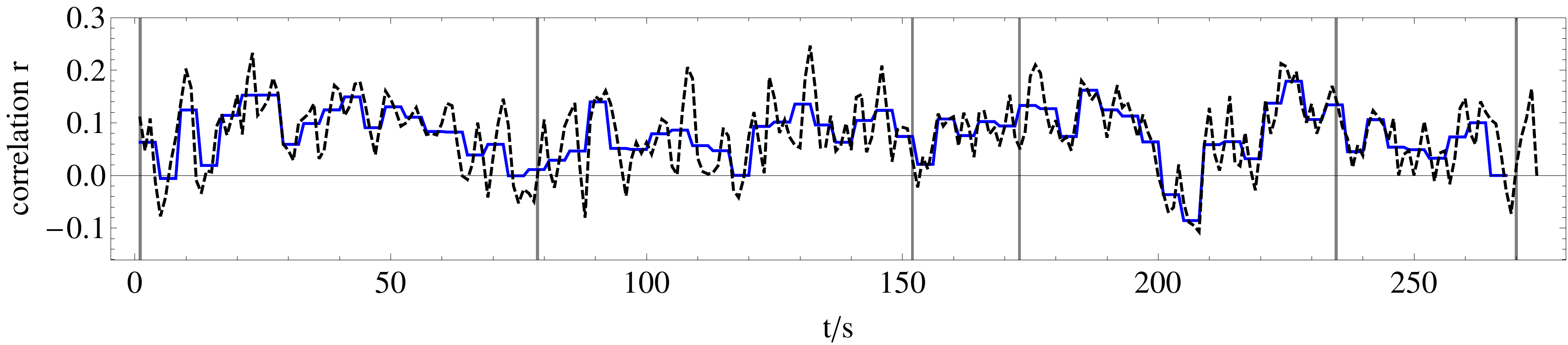}
    \caption{Example of the synchronization dynamics between two electrodes. Dashed black line: Time series of the Pearson correlation coefficient $r$ calculated for successive 1-second time windows ($n = 500$ in Eq.\,\eqref{eq:Pearson}) between averaged EEG recordings of electrode Fp1 (lower plot in Fig.\,\ref{fig.5}) and electrode T7. Blue line: Pearson correlation coefficient averaged over 4 consecutive 1-second time windows of the dashed black line. 
		}
    \label{fig.7}
\end{figure}

For each electrode pair of these nine data sets filtered in this way, the synchronization is calculated in the next step. Thus, we can analyze the synchronization dynamics as a function of the frequency bands. As a synchronization measure the linear cross-correlation, for simplicity taken without time delays, is used, which in fact results in the Pearson correlation coefficient $r$. This is widely used as a non-directed measure of the strength of the correlation between two variables or sequences $\{x_1,x_2,\dots,x_n\}$ and $\{y_1,y_2,\dots,y_n\}$ \citep{GLA02,BAS15c,ERR17}:

\begin{equation}
  r = r_{x,y} = \frac{\tfrac{1}{n}\sum_{i=1}^{n}(x_i-\overline{x})(y_i-\overline{y})}{\sqrt{\tfrac{1}{n}\sum_{i=1}^{n}(x_i-\overline{x})^2}\sqrt{\tfrac{1}{n}\sum_{i=1}^{n}(y_i-\overline{y})^2}},
\label{eq:Pearson}
\end{equation}
where $\overline{x},\overline{y}$ denotes the mean of $x,y$, respectively. Since we aim to reveal synchronization dynamics on the level of musical form, we calculate the correlation within successive 1-second time windows for each possible pair of electrodes of each wavelet-filtered dataset, which results in 32*31/2*9 = 4464 time series with a resolution of 1 second, and each time series has a length of 270 s corresponding to the stimulus length (see Fig.\,\ref{fig.7}). In recent decades, various methods for measuring synchronization have been introduced \citep{BAS15c,BLI11}. The advantage of the Pearson correlation coefficient $r$ is that it allows for easy and efficient calculation of the linear correlations between two variables or time series, and the results are very similar to those obtained by other common methods such as the phase-locking value \citep{LAC99}. For a comparison of the different synchronization measures see \citep{JAL14b}. 

\begin{figure}
\centering
\includegraphics[width = .9\textwidth]{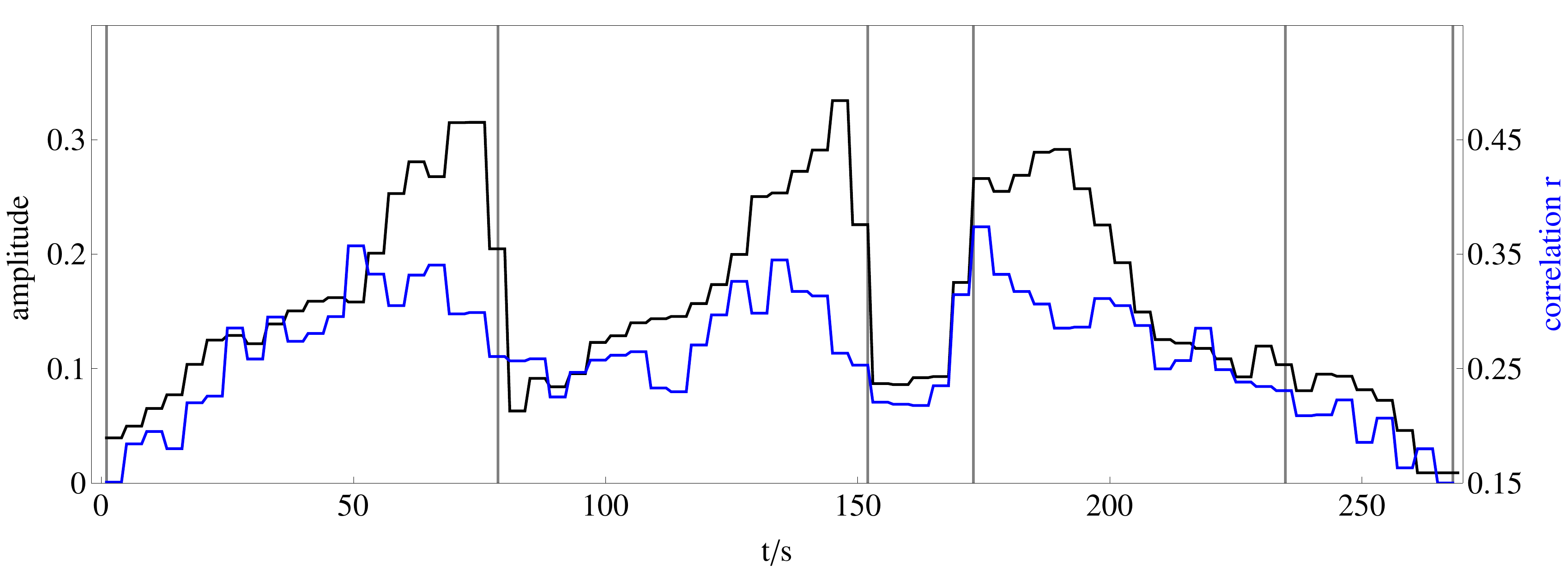}
    \caption{Comparison of whole brain synchronization dynamics and representation of the musical form of the stimulus. The black line shows the amplitude dynamics of the stimulus as a representation of the musical form, averaged over each of 4 consecutive seconds. The blue line shows the average of the 25 correlation time series between two electrodes from each frequency band that correlates most strongly with the amplitude dynamics of the stimulus.}
    \label{fig.8}
\end{figure}

\begin{figure}
\centering
\includegraphics[width = .9\textwidth]{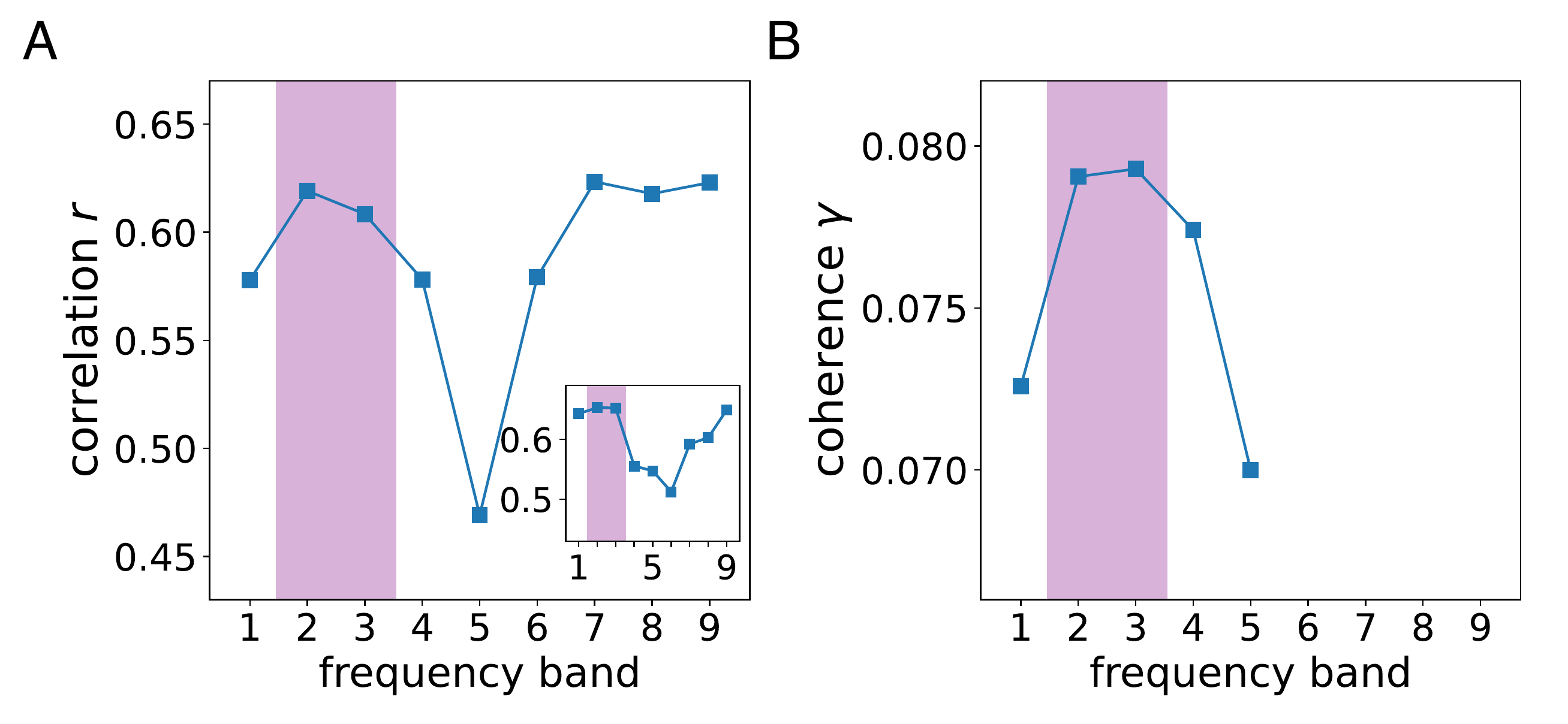}
    \caption{Comparison between experimental and numerical results: (A) Experimentally recorded correlation $r$ of the individual averages of the amplitude time series for each frequency band most strongly correlated with the stimulus as a function of frequency band (FB) FB 1: $125 - 250$ Hz, FB 2: $62.5 - 125$ Hz, FB 3: $31.25 - 62.5$ Hz, FB 4: $15.63 - 31.25$ Hz, FB 5: $7.81 - 15.63$ Hz, FB 6: $3.91 - 7.81$ Hz, FB 7: $1.95 - 3.91$ Hz, FB 8: $0.98 - 1.95$ Hz, FB 9: $0.49 – 0.98$ Hz. The inset depicts the Pearson correlation coefficient $r$ as a function of frequency band where instead of the amplitude the fractal dimension \citep{GRA83,GRA83a} has been used for the calculation of $r$. (B) Numerically simulated coherence $\gamma$ between network dynamics and external stimulus, where the corresponding frequency bands are averaged from Fig.\,\ref{fig.3}. As in Fig.\,\ref{fig.3}, the purple shaded regions in both panels indicate the gamma-band ($f_b \approx 30-120$ Hz), respectively.
		}
    \label{fig.9}
\end{figure}

In order to relate this huge number of time series to the amplitude modulation of the stimulus, we first average the amplitude of the stimulus and the correlation coefficients calculated for the 496 electrode pairs and 9 frequency bands within successive 4-second windows to avoid minor amplitude fluctuations and obtain a scaling corresponding to about two musical bars that fits to changes related to the musical form (Fig.\,\ref{fig.7}). In the second step, we correlate all 4464 time series from the synchronization analysis with the amplitude time series of the stimulus. In the third step, we select the 25 time series per frequency band that correlate most strongly with the time series of the amplitude of the stimulus, shown in Fig.\,\ref{fig.8}. Now, we average these 25 time series per frequency band, which results in a single time series of 270 s length for each frequency band, respectively. These averaged time series for each  frequency band are correlated over the whole recorded time with the amplitude dynamics of the stimulus (see Fig.\,\ref{fig.9}A). In can be shown that the low and the high gamma-band (FB 2 and FB 3) correlate strongly with the stimulus as expected, but also the slow oscillations (FB 7-9) correlate very well (see discussion below). By this, we can reveal how good the synchronization dynamics in each frequency band corresponds to the amplitude dynamics of the stimulus on the level of musical form. In the next step, we average these time series over all frequency bands and correlate the resulting time series with the amplitude dynamics of the stimulus as well. These two time series correlate with a Pearson coefficient of $r = 0.76$. Therefore, we can conclude that the higher the amplitude of the stimulus, the higher the synchronization between the most correlated time series of the different frequency bands. According to \citep{COH92}, this is a strong effect. 

Note that the correlation between sound amplitude (perceived loudness) or other parameters like brightness or fractal correlation dimension (see inset of Fig.\,\ref{fig.9}A) and brain synchronization is not trivial. First, brain synchronization appears at frequencies much lower than most musical frequencies. Secondly, synchronization appears with multiple perceptual parameters. Thirdly, increasing, e.g., the sound amplitude might lead to an increase of the network amplitude, but here it leads to an enhanced synchronization, pointing to a highly nonlinear process in the network, caused by the activity of the brain when perceiving sound.

It is interesting to note that the correlation with the stimulus is highest when the time series from all frequency bands are averaged. The correlation coefficient of the averages of the 25 most correlated time-series as a function of the individual frequency bands is shown in Fig.\,\ref{fig.9}A. It shows two regimes of high correlation, separated by a frequency band (FB 5) with low correlation. Here, the central nervous system in the spinal cord and its relation to the locomotor system are expected to be responsible for the dynamics in the frequency bands FB 6-9 due to their frequency range close to walking and dancing \citep{NOO99}. Note that the electroencephalogram (EEG) recordings are performed on the skull, and therefore represent the brain dynamics of the neocortex which is interacting with the brain stem. Therefore, the high correlations between synchronization and musical form in bands FB 6-9 can be interpreted as caused by the interaction of the neocortex with subcortical brain regions. Likewise, the high correlations in bands FB 2-3 are interpreted as activity of the neocortex solely, as expected. The results therefore also suggest a separation of musical form-related synchronization between cortical (bands 2-3) and subcortical (bands 6-9) regions.

The high correlations observed in frequency bands 2 and 3 for the sound amplitude (see Fig.\,\ref{fig.9}A) as well as for the fractal correlation dimension (see inset of Fig.\,\ref{fig.9}A) correspond to a frequency range of $31.25-125$ Hz (gamma-band). On the other hand in Fig.\,\ref{fig.3}, the strongest coherence between the Kuramoto order parameter (measure for global neural synchronization) and the external input can be found for $n_b=10-40$ Hz. Taking into account that the natural frequency of each node is $f_{\text{FHN}} \approx 0.4$, we can calculate the corresponding frequency band $f_b = n_b/f_{\text{FHN}}$. As shown by the upper x-axis in Fig.\,\ref{fig.3}, the strongest coherence in our model can be observed for a frequency band of $f_b = 40-100$ Hz, which agrees with the gamma-band in the brain. For comparison with the experiment, we show the corresponding numerically simulated results in Fig.\,\ref{fig.9}B, where the respective frequency bands are averaged from Fig.\,\ref{fig.3}. Both experimental and numerical results show a pronounced maximum of correlation between stimulus and brain dynamics for the gamma-band (bands 2 and 3) in Fig.\,\ref{fig.9}. Note that the second maximum in the experimental data (panel A), which is due to the interaction of the neocortex with subcortical brain regions as discussed above, is absent in the simulated data (panel B) since the computer simulation is only performed for the neocortex, using a cochlea input, but neglecting brain stem activity. 

\section{Conclusion}

We have investigated the influence of music in a simulated network of FitzHugh-Nagumo oscillators with empirical structural connectivity obtained from healthy human subjects, and have compared it to measured electroencephalogram (EEG) data. We report an increase of coherence between the global dynamics and the input signal induced by a specific music song. We have shown that the level of coherence depends on the frequency band. We have compared our results with experimental data, which describe global neural synchronization between different brain regions in the gamma-band range and its increase just before transitions between different parts of the musical form (musical high-level events). Such synchronization increases before musical large-scale form boundaries, and decreases afterwards, therefore represents musical large-scale form perception.

The transformation of sound into neural spikes takes place in the cochlea, a part of the human ear which is directly connected to the auditory cortex. By means of the basilar membrane, the brain is able to perceive different frequencies organized in so-called critical bands. We have applied a cochlea model to transform a specific music song into an input signal representing neural spikes evoked by the music song. This input signal has then been supplied to a simulated network of neural oscillators with empirical structural connectivity. By the transformation of the dimensionless time units of the oscillator model to real time units, we have investigated dynamical scenarios in dependence on the introduced frequency band parameter. To quantify moreover the overlap between input signal and network dynamics, we have introduced a coherence measure. It has turned out that this coherence measure depends sensitively on the frequency band and has its maximum in the gamma-band. Therefore, depending on the frequency band, coherence can be induced between the dynamics of the system and its input signal.

These results are in accordance with our own and previous experiments \citep{HAR14, HAR20a} where music has also been found to induce a certain degree of synchrony in the human brain. We have shown that listening to music can have a remarkable influence on the brain dynamics, in particular, a periodic alternation between synchronization and desynchronization which is strongly related to the music perceived. We have experimentally analyzed in detail the influence of real music on the neural activity with respect to the common frequency bands in the brain. By means of the Pearson correlation coefficient of the sound amplitude as well as the fractal correlation dimension, we have found the gamma-band to be important for musical form perception. Just as in the computer simulation, we have found a pronounced maximum for this frequency range. Furthermore, the results suggest a separation in musical form-related brain synchronization between high brain frequencies, associated with neocortical activity, and low frequencies in the range of dance movements, associated with interactivity between cortical and subcortical regions. Besides, an alternation between synchronization and desynchronization reflects the variability of the system; this can be seen as a critical state between a fully synchronized and a desynchronized state. It is known that the brain is operating in a critical state at the edge of different dynamical regimes \citep{MAS15a,SHI22}, exhibiting hysteresis and avalanche phenomena as seen in critical phenomena and phase transitions \citep{KIM18,RIB10,STE10a}.

By choosing appropriate parameters and measures, we have reported an intriguing dynamical behavior in dependence on the frequency bands, and have observed the induced increase of coherence both in numerical and experimental setups. To sum up, music supplied to the brain allows for a high coherence and correlation between musical input and brain dynamics especially in the gamma-band. This insight may be used to fathom the general modalities of the influence of music on the human brain.

\section*{Conflict of Interest Statement}

The authors declare that the research was conducted in the absence of any commercial or financial relationships that could be construed as a potential conflict of interest.

\section*{Author Contributions}

JS did the numerical simulations and the theoretical analysis, LH has performed the experiments. RB and ES supervised the study. All authors designed the study and contributed to the preparation of the manuscript. All the authors have read and approved the final manuscript.

\section*{Funding}
This work was supported by the Deutsche Forschungsgemeinschaft (DFG, German Research Foundation, project No.\,429685422) and the Open Access Publication Fund of TU Berlin.

\section*{Acknowledgments}
We are grateful to Anton\'{i}n \v{S}koch and Jaroslav Hlinka for preparing the example structural connectivity matrices.

%

\bibliographystyle{frontiersinSCNS_ENG_HUMS} 



\end{document}